\documentclass[11pt,twoside]{article}
\usepackage[affil-it]{authblk}
\usepackage[a4paper,top=2.5cm,left=3cm,right=3cm,bottom=2.5cm]{geometry}
\usepackage{amsmath}
\usepackage{multirow}
\usepackage{booktabs}
\usepackage[ruled,vlined]{algorithm2e}
\usepackage{array}
\usepackage{color}
\usepackage[normalem]{ulem}
\usepackage{hyperref}
\usepackage{amssymb}
\usepackage{graphicx}
\usepackage{amsfonts}
\usepackage[english]{babel}
\usepackage{natbib}

\begin{document}

\title{GPU Accelerated Image Quality Assessment-Based Software for Transient Detection}

\author{X. Li, K. Ad\'{a}mek, and W. Armour}
\affil{University of Oxford, Oxford, the United Kingdom; \texttt{wes.armour@oerc.ox.ac.uk}}
\date{}

\maketitle

\begin{abstract}
Fast imaging localises celestial transients using source finders in the image domain. The need for high computational throughput in this process is driven by next-generation telescopes such as Square Kilometre Array (SKA), which, upon completion, will be the world's largest aperture synthesis radio telescope. It will collect data at unprecedented velocity and volume. Due to the vast amounts of data the SKA will produce, current source finders based on source extraction may be inefficient in a wide-field search. In this paper, we focus on the software development of GPU-accelerated transient finders based on Image Quality Assessment (IQA) methods --- Low-Information Similarity Index (LISI) and augmented LISI (augLISI). We accelerate the algorithms using GPUs, achieving kernel time of approximately 0.1 milliseconds for transient finding in $2048 \times 2048$ images.
\end{abstract}

\section{Introduction}

Fast imaging localises celestial transients by using source detection algorithms in radio astronomical images. These transients are usually concentrated in a small section of the image in wide-field observation, while the rest of the image remains largely stable with slight variations over time. Since it can be challenging for human visual system to identify transients within these images, we develop transient finders based on two intensity-sensitive Image Quality Assessment (IQA) methods: Low-Information Similarity Index \citep[LISI;][]{Wider} and augmented LISI \citep[augLISI;][]{iqara}, as expressed by
\begin{equation}
\label{LISI}
\mathrm{LISI}\left( {\mathbf{x},\mathbf{y}} \right) = D\frac{\sum\limits_{i=1}^{N}\frac{\left| {x_{i} + y_{i}} \right|}{\left| {x_{i} - y_{i}} \right| + C_{1}}}{\max\left( {\sum\limits_{i=1}^{N}x_{i}},{\sum\limits_{i=1}^{N}y_{i}} \right) + C_{2}},
\end{equation}
\begin{equation}
    \label{augLISI}
    \mathrm{augLISI}\left( {\mathbf{x},\mathbf{y}} \right) = 1 - \frac{\sum\limits_{i=1}^{N}\left| {x_{i} + y_{i}} \right| \left| {x_{i} - y_{i}} \right|}{\sum\limits_{i=1}^{N}x_{i}+\sum\limits_{i=1}^{N}y_{i}+ C},
\end{equation} 
where $\mathbf{x}$ and $\mathbf{y}$ are input images, with $C_{1}<<1$, $C_{2}<<1$, and $C<<1$. Both IQA methods yield values from 0 to 1, with 1 indicating identical input images. LISI works best for images with only point sources and minimal noise, whereas augLISI is better suited for images containing extended sources or higher levels of noise.

\section{Methodology}

To enhance the efficiency, we leverage GPUs to accelerate algorithms through CUDA programming and parallel computing, and implement GPU-accelerated transient finders based on LISI \footnote{\url{https://github.com/egbdfX/gpuLISI}} and augLISI \footnote{\url{https://github.com/egbdfX/gpuAugLISI}}. A C/C++ interface ensures portability, allowing input of FITS images, passing arrays to CUDA functions, and outputting the corresponding IQA matrix. During pre-processing, negative pixel values in the inputs are set to zeros. The input images are divided into tiles for transient localisation. The key CUDA kernels, \texttt{split\_lisi} and \texttt{split\_auglisi}, compute LISI and augLISI values for each tile in parallel, where image and tile sizes are $N \times N$ and $N_T \times N_T$ pixels, respectively.

According to Equation (\ref{LISI}) and (\ref{augLISI}), the pixel-to-pixel summation is the most computationally expensive operation. Each tile is processed by a thread block, with computations remaining independent at both the tile and pixel levels. We use a reduction algorithm for efficient aggregation, leveraging shared memory for intermediate results. Threads process batches of pixels to compute partial sums, which are further reduced to obtain final LISI or augLISI values.

\begin{figure}[!ht]
    \centering
    \includegraphics[width=0.48\textwidth]{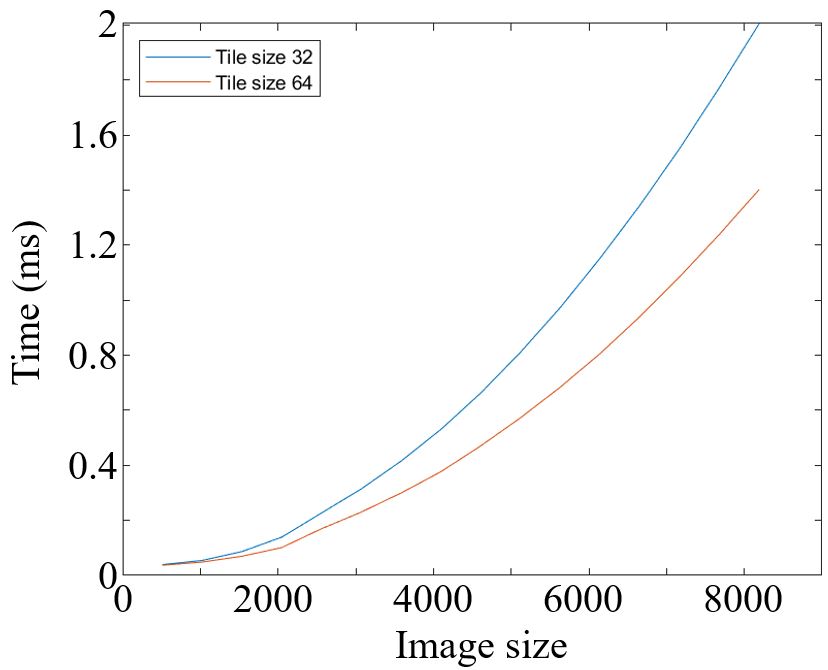}
    \hfill
    \includegraphics[width=0.48\textwidth]{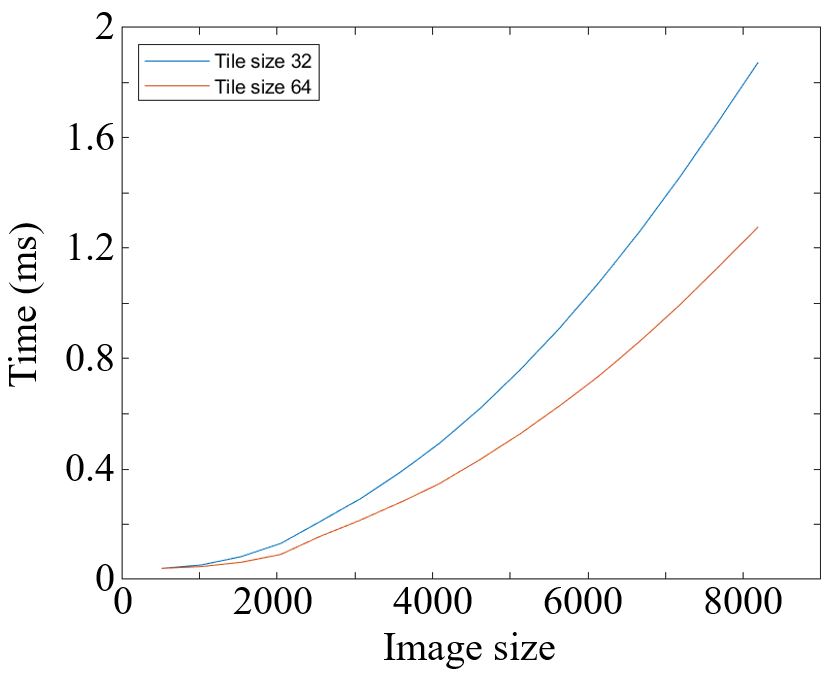}
    \caption{Kernel execution time for GPU-accelerated transient finders based on \emph{Left:} LISI and \emph{Right:} augLISI.
\label{extratimechap3}}
\end{figure}

On an NVIDIA H100 GPU \footnote{\url{https://resources.nvidia.com/en-us-tensor-core}}, the kernels achieve median execution times of 0.101 ms (LISI) and 0.091 ms (augLISI) for $N=2048$ and $N_T=64$ over 10 runs. Kernel execution times for different configurations are shown in Fig. \ref{extratimechap3}.

\section{Results}

\begin{figure}[!ht]
    \centering
    \includegraphics[width=0.32\textwidth]{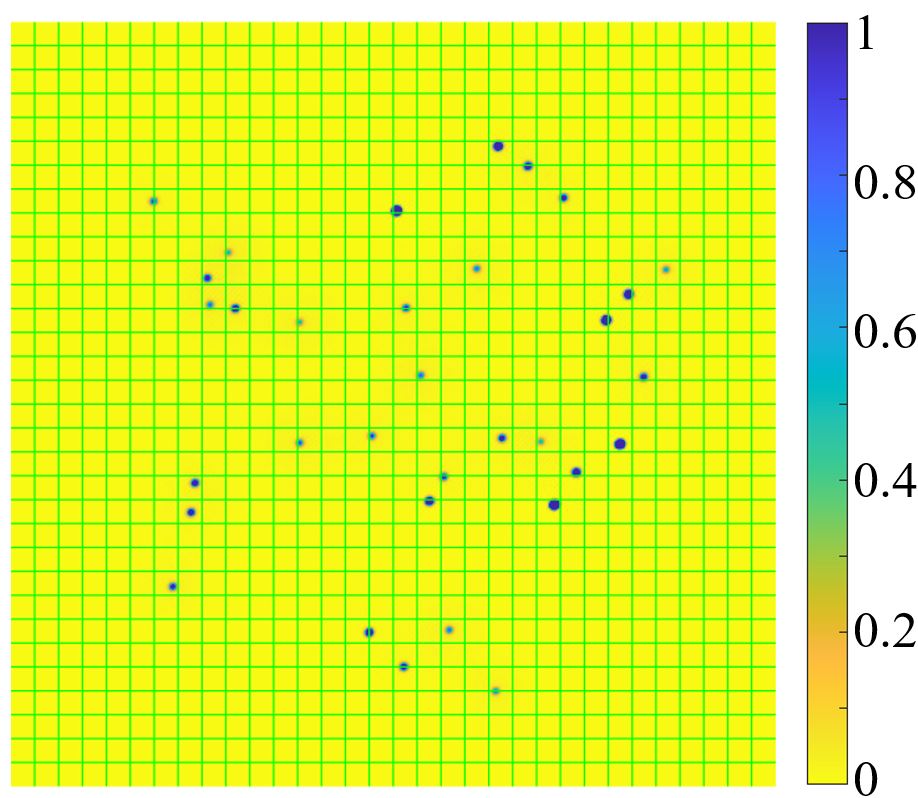}
    \hfill
    \includegraphics[width=0.32\textwidth]{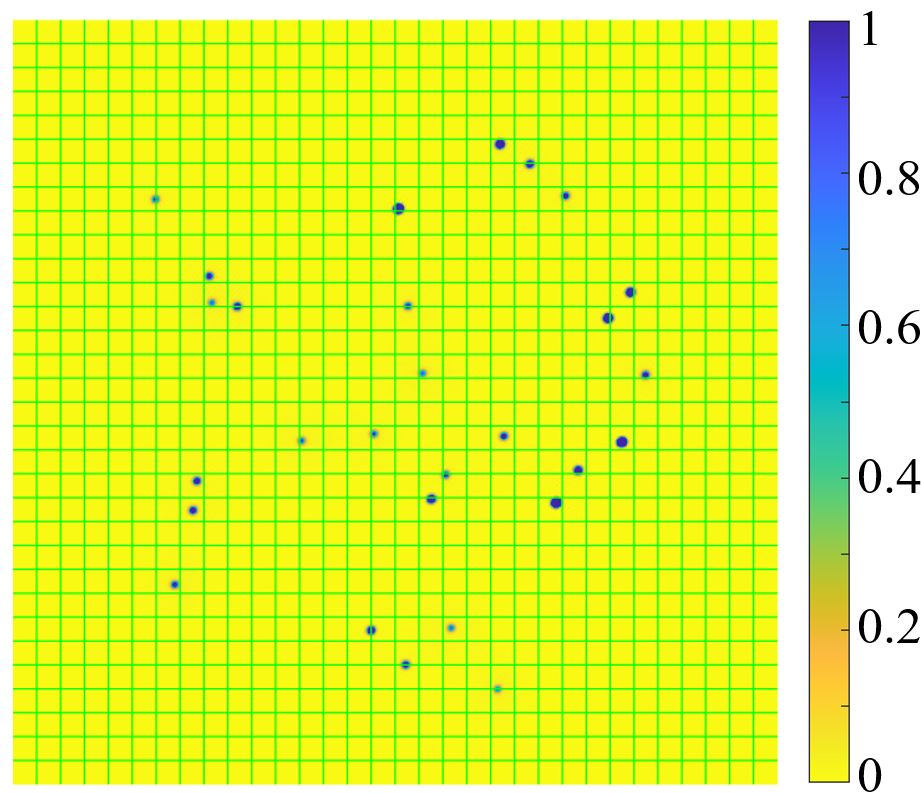}
    \hfill
    \includegraphics[width=0.32\textwidth]{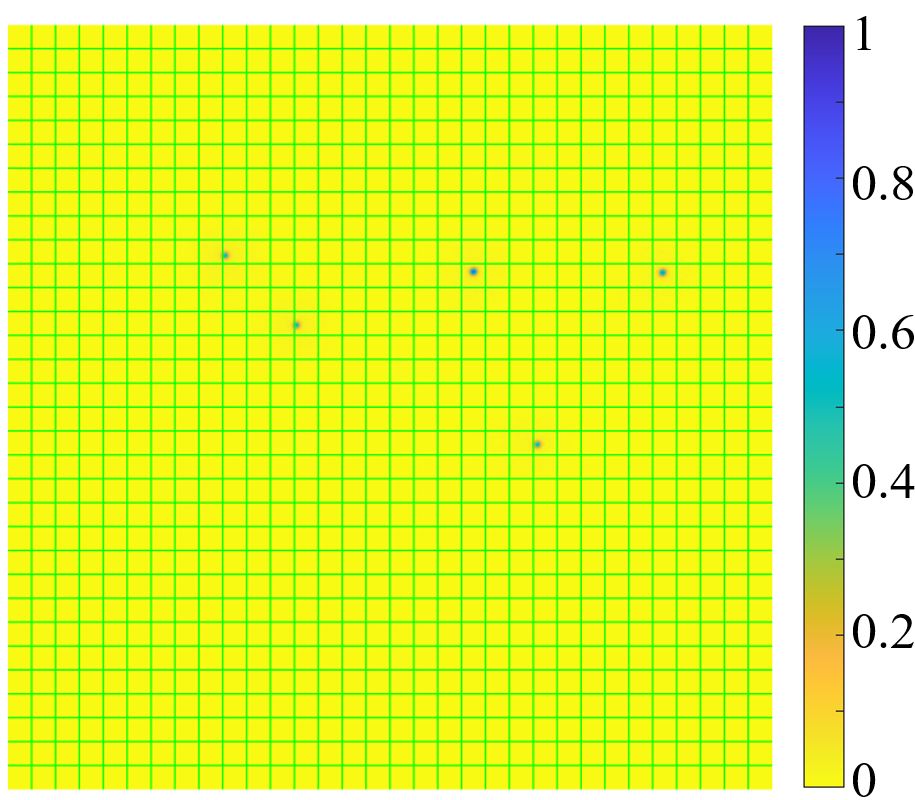}
    \caption{Simulated radio astronomical images. \emph{Left:} Restored image of MS1. \emph{Middle:} Restored image of MS2. \emph{Right:} Difference image between the two restored images, highlighting the changes, with five sources visible.
\label{iqatile}}
\end{figure}

To evaluate our transient finders, we simulate Measurement Sets (MS) using source information included in GaLactic and Extragalactic All-sky MWA (GLEAM) catalogue \citep{gleam} with SKA1-LOW telescope configuration, employing Oxford's Square Kilometre Array Radio-telescope simulator \citep[OSKAR;][]{OSKAR1}. Images are restored using Hogbom CLEAN \citep{CLEAN1}, as shown in Fig. \ref{iqatile}. The first dataset (MS1) contains 33 sources, while the second (MS2) excludes 5 low-intensity sources from MS1 to mimic transient disappearance. Transients typically exhibit on-and-off emission behaviour, causing intensity variations between observations of the same section of the sky, while fixed sources remain unchanged. To identify and characterise these changes, the images are divided into tiles for position-based similarity analysis. Here, the image contains $2048 \times 2048$ pixels, with each tile sized at $64 \times 64$ pixels.

By applying our transient finders to the images of MS1 and MS2, we compute tile-wise similarities, yielding an IQA (LISI or augLISI) matrix. The LISI or augLISI values for individual matrix elements, each representing the similarity of a tile in the images, can be aggregated into a single value equivalent to the LISI or augLISI of the similarity of the entire images, as expressed by
\begin{equation}
\label{splitLISI}
\mathrm{LISI}\left( {\mathbf{x},\mathbf{y}}\right) = \frac{\sum\limits_{t=1}^{n}\left\{\mathrm{lisi}_t \times \left[\mathrm{max}\left(\sum\limits_{k=1}^{N_T \times N_T}x_k,\sum\limits_{k=1}^{N_T \times N_T}y_k\right)+C_2\right]\right\}}{\mathrm{max}\left(\sum\limits_{i=1}^{N \times N}x_{i},\sum\limits_{i=1}^{N \times N}y_{i}\right)+ C_2},
\end{equation}
\begin{equation}
    \label{splitaugLISI}
    \mathrm{augLISI}\left( {\mathbf{x},\mathbf{y}} \right) = \frac{\sum\limits_{t=1}^{n}\left[\mathrm{auglisi}_t \times \left(\sum\limits_{k=1}^{N_T \times N_T}x_k+\sum\limits_{k=1}^{N_T \times N_T}y_k+C\right)\right]-\left(n-1\right)C}{\sum\limits_{i=1}^{N \times N}x_{i}+\sum\limits_{i=1}^{N \times N}y_{i}+ C}.
\end{equation} 

Here, $k$ ($k = 1,2,...,N_T \times N_T$) denotes a pixel index within a tile, and $i$ ($i= 1,2,...,N \times N$) denotes a pixel index in the entire image. The image contains $n$ tiles, with:
\begin{equation*}
    n = \frac{N}{N_T} \times \frac{N}{N_T}.
\end{equation*}
In Equation (\ref{splitLISI}), $\mathrm{LISI}$ indicates the LISI value for the entire image, and $\mathrm{lisi}_t$ indicates the LISI value for the $t$-th tiles ($t = 1, 2, ..., n$). Similarly, in Equation (\ref{splitaugLISI}), $\mathrm{augLISI}$ refers to the augLISI value for the entire image, and $\mathrm{auglisi}_t$ refers to the augLISI value for the $t$-th tiles ($t = 1, 2, ..., n$). Notably, the ``entire image'' may also refer to a larger tile formed by merging smaller ones.

\begin{figure}[!ht]
    \centering
    \includegraphics[width=0.48\textwidth]{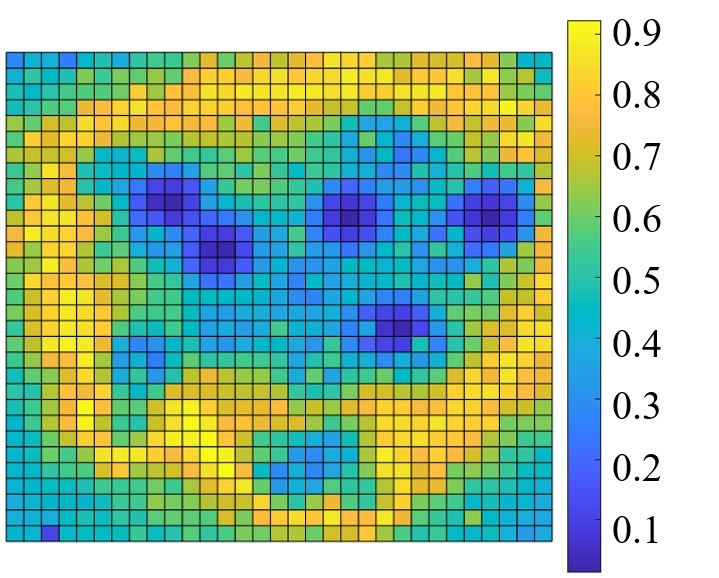}
    \hfill
    \includegraphics[width=0.48\textwidth]{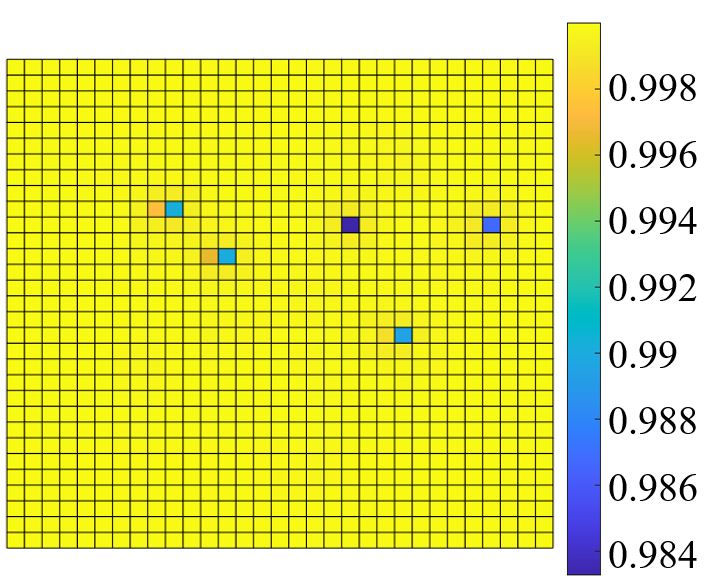}
    \caption{IQA results comparing corresponding tiles of the MS1 and MS2 images. \emph{Left:} LISI results. \emph{Right:} augLISI results.
\label{tilevalue}}
\end{figure}

The LISI and augLISI results comparing corresponding tiles of the MS1 and MS2 images are illustrated in Fig. \ref{tilevalue}, with colour bars indicating the IQA values.

Figure \ref{tilevalue} highlights the strengths and limitations of each approach. LISI is highly sensitive to changes across all intensity levels, whether high or low, even when the variations are subtle. This sensitivity is reflected in the wide range of LISI values across the image, including low values in the corners caused by the CLEAN algorithm and spread structures from the CLEAN beams which convolved on the model image. On the other hand, augLISI focuses on significant changes, with reduced sensitivity to minor variations, as shown by the narrower range of augLISI values. It effectively identifies source differences while disregarding noise variations, localising transients to tiles with fewer false positives. 

\section{Conclusion}

GPU-accelerated transient finders, based on LISI and augLISI, are designed for transient detection in radio astronomical images. These transient finders exhibit both high sensitivity and efficiency in identifying and localising sources with variations. While LISI achieves greater sensitivity than augLISI, it is also more prone to false positives. In contrast, augLISI delivers higher accuracy for transient localisation.

\section*{Acknowledgements}
The authors are grateful to Fred Dulwich and Ben Mort for their valuable guidance on OSKAR.

\bibliographystyle{plainnat}
\bibliography{P612.bib}

\end{document}